\documentclass[a4paper]{jpconf}
\usepackage{graphicx}
\usepackage{epsfig}
\usepackage{dcolumn}
\usepackage{bm}
\newcommand{\comment}[1]{}

\def\beq{\begin{equation}}
\def\eeq{\end{equation}}
\def\bea{\begin{eqnarray}}
\def\eea{\end{eqnarray}}
\def\rpv{\not\!\!{R_P}}
\begin{document}
\title{Neutrino masses and mixing in $\mu\nu$SSM}

\author{Pradipta Ghosh}

\address{Department of Theoretical Physics and Centre for Theoretical
Sciences,\\
 Indian Association for the Cultivation of Science,\\
 2A $\&$ 2B Raja
S.C. Mullick Road,\\
 Kolkata 700 032, India}

\ead{tppg@iacs.res.in}
\begin{abstract}
$\mu\nu$SSM is an $R$-parity violating non-minimal supersymmetric model
which uses right chiral neutrino superfields to solve the $\mu$-problem.
The $R$-parity violation together with a TeV scale seesaw mechanism using right
handed neutrinos are instrumental for the light neutrino mass generation
in $\mu\nu$SSM. We show that it is possible to accommodate three flavour
global neutrino data in $\mu\nu$SSM  with three massive neutrinos at the 
tree level. Ingression of the one-loop corrections to neutrino masses and 
mixing shows certain variations over the tree level analysis depending on the
specific hierarchy of neutrino masses involved. In $\mu\nu$SSM some of
the $R$-parity violating decay branching ratios of the lightest neutralino 
show nice correlation with certain neutrino mixing angle. These correlations
along with the presence of displaced vertices in the decay of the lightest
neutralino can be further investigated as a test of $\mu\nu$SSM in collider experiments.
\end{abstract}
\section{Introduction}
 Recent findings of neutrino oscillation experiments has put forward
strong evidences for the massive neutrinos. A satisfactory 
explanation for the non-zero neutrino masses is beyond the Standard Model framework.
Weak scale supersymmetry (SUSY) is a well motivated candidate for beyond 
the Standard Model (SM) physics and is of immense interest with the initiation
of the Large Hadron Collider (LHC) era. However, SUSY has its own theoretical
problems out of which a well-known name is the $\mu$-problem \cite{Kim-Nilles}.
Solutions to the $\mu$-problem are addressed in a non-minimal version
of SUSY model using a SM gauge singlet superfield $(\hat S)$.  On the other hand,
both the minimal supersymmetric standard model (MSSM) and next-to-minimal 
supersymmetric standard model (NMSSM) predicts massless neutrinos
similar to the SM. Massive neutrinos in a SUSY framework can be achieved in two
ways, either using the $R$-parity violation ($\rpv$) or through canonical 
seesaw mechanism (initially proposed to prevent fast proton decay through sparticle
mediated processes). $R_p$ is defined as  $R_p = (-1)^{L+3B+2s}$ with
$L(B)$ as lepton(baryon) number and $s$ is the spin. MSSM can accommodate 
neutrino masses through the bilinear or trilinear $\rpv$ without solving 
the $\mu$-problem. However, these new bilinear $\rpv$ violating terms
will again cause {\it{naturalness}} problem similar to the $\mu$-problem 
\cite{Nilles-Polonsky}.

\vspace{0.2cm}
The ``$\mu$ from $\nu$'' supersymmetric standard model ($\mu\nu$SSM) 
\cite{munoz-lopez_fogliani,munoz-lopez_fogliani-2} invokes right-handed
neutrino superfields to solve the $\mu$-problem and at the same time 
uses the same set of right-handed
neutrino superfields to generate light neutrino masses. In a nutshell
$\mu\nu$SSM is a {\it{minimalistic extension}} of MSSM (includes only
right-handed neutrino superfields apart from the MSSM superfields) for
accommodating neutrino masses and simultaneously solving the $\mu$-problem.
 
\vspace{0.2cm}
The $\mu\nu$SSM can fit the three flavour global neutrino data
even with flavour diagonal structure of the neutrino Yukawa couplings
for various hierarchies of light neutrino masses \cite{pg1} at the tree level. 
All three light neutrinos can acquire non-zero masses at the tree level. 
$\rpv$ in $\mu\nu$SSM will lead to an unstable Lightest
Supersymmetric Particle (LSP). Some ratios of the decay branching ratios
of the LSP show nice correlations with certain neutrino mixing angles \cite{pg1}
depending on the LSP nature and the hierarchy in the light neutrino
masses. These correlations can be investigated as the experimental signatures
of $\mu\nu$SSM with possible discriminating features from other SUSY models.
Another important feature of $\mu\nu$SSM is the possibility of having
displaced vertices in the decay of the lightest neutralino, which can vary
from a few {\it{mm}} to $\sim$ $1$ {\it{meter}} depending on the nature
of the lightest neutralino \cite{pg1}.
Tree level results of neutrino masses and mixing show variations
with the addition of the one-loop radiative corrections. The amount of variations
were observed to be dependent on the choice of the hierarchy in light neutrino
masses \cite{pg2}. 
 
\vspace{0.2cm}
We note in passing that various other aspects of $\mu\nu$SSM like
LHC phenomenology, spontaneous CP-violation, gravitino dark matter,
baryogenesis are discussed in refs.\cite{Porod-Bartl,munoz-lopez-3,
munoz-lopez-4,Chung} respectively. For a review of $\mu\nu$SSM and seesaw mechanism in
$\mu\nu$SSM see refs. \cite{munoz-munussm} and \cite{lopez-seesaw}.

\section{The model}
The model superpotential is

\bea 
W &=& \epsilon_{ab}(Y^{ij}_u\hat H^b_2\hat Q^a_i\hat u^c_j +
Y^{ij}_d\hat H^a_1 \hat Q^b_i\hat d^c_j + Y^{ij}_e\hat H^a_1\hat
L^b_i\hat e^c_j + Y^{ij}_\nu
\hat H^b_2\hat L^a_i\hat \nu^c_j)\nonumber \\
&-&\epsilon_{ab} \lambda^i\hat \nu^c_i\hat H^a_1\hat H^b_2 +
\frac{1}{3}\kappa^{ijk}\hat \nu^c_i\hat \nu^c_j\hat \nu^c_k,
\label{superpotential}
\eea
where $\hat \nu^c_i$  are the right chiral neutrino superfields
ingressed in $\mu\nu$SSM apart from the MSSM $SU(2)$ doublet 
($\hat H_1,\hat H_2,\hat Q_i,\hat L_i$)
and singlet (${\hat u}^c_j,{\hat d}^c_j,{\hat e}^c_j$) superfields.
In eq.(\ref{superpotential}) $i,j,k$ represent generational indices.
Appearance of any bilinear term is prohibited by imposing a
$Z_3$ symmetry in the $\mu\nu$SSM superpotential.
This superpotential explicitly breaks $R_p$ through lepton number
violation by odd units ($5^{th}$ and $6^{th}$ terms of eq.(\ref{superpotential})).
There are corresponding entries in the soft SUSY breaking sector 
too \cite{munoz-lopez_fogliani}.

\vspace{0.2cm}
After the electroweak symmetry breaking (EWSB) when the scalar component
of Higgses and left and right sneutrino fields develope the respective
Vacuum Expectation Values (VEVs) $~~v_1, v_2, v'_i, v^c_i$, we have an effective
$\mu$-term and bilinear $\rpv$ terms ($\epsilon_{i}$) as
$\mu = \sum{\lambda^i}{v^c_i}$ and $\epsilon^i = \sum Y^{ij}_\nu v^c_j$, respectively.
The $6^{th}$ term of eq.(\ref{superpotential}), with the coefficient $\kappa^{ijk}$, is 
included in order to avoid an unacceptable axion associated to the breaking of a
global $\rm U(1)$ symmetry \cite{ellis-gunion-haber}. After EWSB this term 
generates effective Majorana masses ($m_{\nu^c_{ij}}$) for the 
right-handed neutrinos at the electroweak (EW) scale and are given by $2~\kappa^{ijk} v^c_k$.
These EW scale right neutrinos are further responsible for light neutrino mass 
generation through seesaw, thereby lowering the seesaw-scale within the reach
of a ${\it{TeV}}$ scale collider.

\vspace{0.2cm}
Here, as a digression, let us mention that the
spontaneous breakdown of the $Z_3$ symmetry through right-sneutrino VEV can in
general lead to the formation of domain walls \cite{dw1,dw2,dw3}. The
associated problems can, however, be ameliorated through well-known methods
\cite{dws1,dws2}.

\section{Tree level neutrino masses and mixing}
The lepton number violating interactions in the superpotential and in the soft 
SUSY breaking part of the scalar potential allows mixing between states having zero
lepton number with states having non-zero lepton number. As a consequence
mass matrices in the scalar and fermion sectors get enhanced over their MSSM structures
\cite{munoz-lopez_fogliani-2,pg1}. In the neutral fermion sector
now the four MSSM neutralinos mix with three generations of left and 
right-handed neutrinos and hence the neutralino mass matrix enlarges
to $10\times10$. In the basis ${\Psi^0}^T = \left(\widetilde B^0, \widetilde W_3^0, 
\widetilde H_1^0, \widetilde H_2^0,{\nu^c_i},{\nu_i}\right)$, the
neutralino mass matrix is given by

\beq 
\mathcal{M}_n =
\left(\begin{array}{cc}
    M_{7\times 7} & m_{3\times 7}^T \\
    m_{3\times 7} & 0_{3\times 3}
\end{array}\right).
\label{neutralino-seesaw}
\eeq

The matrix $M_{7\times7}$ contains a $4\times4$ block of MSSM
neutralinos as well as a $3\times3$ block of right-neutrinos and mixing 
terms between them \cite{munoz-lopez_fogliani-2,pg1}. 
The null $3\times3$ block in
$\mathcal{M}_n$ signifies the absence of Majorana mass terms for the left
handed neutrinos. The elements of $m_{3\times7}$ contain either left handed
sneutrino VEVs $(v^{\prime}_i)$ or Higgs VEVs multiplied by neutrino Yukawa 
couplings $(Y_{\nu}^{ij})$, and hence, are of much smaller magnitudes compared 
to the entries of $M_{7\times7}$. This feature ensures a {\it{seesaw}}-like 
structure of $\mathcal{M}_n$\cite{munoz-lopez_fogliani-2,pg1} and the light neutrino
mass matrix using seesaw is given by 

\beq 
{M^{seesaw}} = -{m_{3\times7}} {M_{7\times7}^{-1}}
{m_{3\times7}^T}.
\label{seesaw_formula}
\eeq

This seesaw mass matrix can be diagonalized (with $m_{\nu_i}$ as mass eigenvalues)
using a unitary matrix $U_{PMNS}$ containing neutrino mixing angles
(provided that the charged lepton mass matrix is already in the diagonal form)
as follows 

\beq
U_{PMNS}^T {M^{seesaw}} U_{PMNS} = diag (m_{\nu_1},m_{\nu_2},m_{\nu_3}).
\label{U_pmns}
\eeq

With a few simplifying assumptions \cite{pg1} one can get an approximate analytical
form for the entries of seesaw mass matrix (${M^{seesaw}}$) as

\beq
M^{seesaw}_{ij} = {\frac{2 A {v^c}}{3 \Delta}} {b_{i}} {b_{j}} 
+ {\frac{1}{6 \kappa {v^c}}} {a_{i}} {a_{j}} 
(1-3\delta_{ij}),
\label{mnuij-compact1}
\eeq
where
\bea
\Delta &=& {\lambda}^2 (v^2_1 +v^2_2)^2 + 4 {\lambda} {\kappa} {v^c}^2 
v_1 v_2 - 4 M \lambda A{\mu}, ~\mu = 3 \lambda v^c,
\nonumber \\
A &=& ({\kappa}{v^c}^2 + {\lambda} v_1 v_2),~\frac{1}{M} = \frac{g^2_1}{M_1} 
+\frac{g^2_2}{M_2}, \nonumber \\
a_i &=& Y_{\nu}^{ii} v_2, ~b_i = (Y_{\nu}^{ii} v_1 + 3 {\lambda} {v'_i}),
\label{clarifications}
\eea
with ${i,j,k} = {e,\mu,\tau}$. We have used the fact that $v'_i\ll v_1,v_2$.

The seesaw structure in $\mu\nu$SSM can be well understood 
if we investigate eq.(\ref{mnuij-compact1}) in the following 
limits\cite{pg1}, (i) ${v^c}\rightarrow \infty$ and $v \rightarrow 0~(v^2=(v^2_1 + v^2_2))$
and (ii) $M\rightarrow \infty$.

In the limit (i) we end up with
\beq
\label{gauginoseesaw}
M^{seesaw}_{ij} \approx -{\frac{{v'_i}{v'_j}}{2 M}},
\eeq
and in case (ii) we have
\beq
\label{ordinaryseesaw}
M^{seesaw}_{ij} \approx{\frac{v^2_2}{6 \kappa {v^c}}}{Y^{ii}_{\nu}}
{Y^{jj}_{\nu}}(1-3 {\delta_{ij}}).
\eeq

The form of eq.(\ref{gauginoseesaw}) is associated with the gaugino 
seesaw effect where the gauge coupling$\times$left sneutrino VEV ($v'_i$) acts as
the {\it{Dirac mass}} and the effective gaugino mass ($M$) plays
the role of the {\it{Majorana mass}}\cite{pg2}. In the gaugino seesaw
scenario because of the presence of left sneutrino VEV,
an effective $\Delta L = 2$ Majorana mass term for light neutrinos
is generated from a pair of $\Delta L = 1$ vertex 
involving left-handed neutrino and neutral wino/bino 
($\widetilde B^0/\widetilde W_3^0$).  This is analogous to
Type-III seesaw because of the association of the hypercharge
zero triplet fermion \cite{pg2}. Another feature of gaugino
seesaw is that it generates only one massive neutrino at the tree level.
On the contrary eq.(\ref{ordinaryseesaw}) is associated with the ordinary seesaw 
effect involving right-handed neutrinos. For this case one can get two massive
neutrinos at the tree level. This is the well-known example of Type-I seesaw
using singlet fermion (right-handed neutrino for $\mu\nu$SSM), where
the $\Delta L = 2$ effect is coming through right chiral
neutrino Majorana masses \cite{pg1}.

\vspace{0.2cm}
With a suitable choice of model parameters one can treat the second term
of eq.(\ref{mnuij-compact1}) as a perturbation over the first term. We
would like to emphasize that for most of the parameter choice $a_i\sim b_i$,
hence the relative weight difference is coming from the co-efficient in front 
(${\frac{2 A {v^c}}{3 \Delta}}$ and ${\frac{1}{6 \kappa {v^c}}}$) \cite{pg1}.
Appearance of $b_i b_j$ in the first term of eq.(\ref{mnuij-compact1})
gives only one non-zero neutrino mass $\propto$ $\sum b^2_i$. 
The other two masses emerge due to the effect of ordinary seesaw \cite{pg1}.
Approximate analytical expressions for the masses and mixing for 
normal hierarchical schemes of light neutrino masses were obtained in
ref. \cite{pg1} which show good agreement with the complete numerical 
analysis without any approximation \cite{pg1}.
As an illustrative example, in case of normal hierarchy the atmospheric mixing 
angle is given by

\bea
\sin^2\theta_{23} = \frac{b^2_\mu}{b^2_\mu + b^2_\tau},
\label{atmos_analytical}
\eea
with $b_i$'s are given by eq.(\ref{clarifications}). It is clear from the
analytical formula in eq.(\ref{atmos_analytical}), that the maximal mixing 
in the atmospheric sector indicates $b_\mu = b_\tau$.

Note that using $b_i = (a_i\cot\beta + 3 \lambda c_i)$ with $c_i = {v'_i}$
and $\rm{tan}\beta = \frac{v_2}{v_1}$,
one ends up with more elucidate form of eq.(\ref{mnuij-compact1}) as \cite{pg2}

\beq
({M^{seesaw}_{\nu}})_{ij} = f_1 a_i a_j + f_2 c_i c_j + f_3 (a_i c_j + a_j c_i),
\label{mnuij-compact-recasted}
\eeq
where (using eq.(\ref{clarifications}))
\bea
f_1 &=& \frac{1}{6 \kappa v^c} (1-3\delta_{ij}) + \frac{2 A v^c {\rm{cot}}^2\beta}{3 \Delta},
~f_2 = \frac{2 A \lambda \mu}{\Delta},~~f_3 = \frac{2 A \mu {\rm{cot}}\beta}{3 \Delta}.
\label{specifications-3}
\eea

It is very clear from eq.(\ref{mnuij-compact-recasted}), that the $1^{st}$
and the $2^{nd}$ term can contribute to only one mass eigenvalue and are
$\propto$ $\sum a^2_i$ and $\sum c^2_i$ with suitable co-efficient in front.
It is the $3^{rd}$ term or the mixing term which is responsible for giving masses
to all three light neutrinos \cite{pg2}.

\section{One-loop corrected neutrino masses and mixing}
It is clear from the above discussion that in $\mu\nu$SSM all three light neutrinos 
get seesaw masses at the tree level even with 
the flavour diagonal choice of neutrino Yukawa couplings ($Y_\nu$) \cite{pg1} consistent 
with the three flavour global neutrino data \cite{Schwetz-Valle}. With the inclusion of
one-loop radiative corrections tree level results of neutrino masses and mixing
receive corrections over their tree level values depending on the concerned
mass hierarchy \cite{pg2}.

\vspace{0.2cm}
There are seven possible sources \cite{pg2} of one-loop corrections to the 
light neutrino masses in $\mu\nu$SSM and they are listed below 
\begin{enumerate}
 \item neutralino - neutral scalar in the loop,
 \item neutralino - neutral pseudoscalar in the loop,
 \item neutralino - $Z$-boson in the loop,
 \item chargino - charged scalar in the loop,
 \item chargino - $W^{\pm}$-boson in the loop,
 \item up-type quark - up-type suqark in the loop,
 \item down-type quark - down-type suqark in the loop.
\end{enumerate}
The relevant Feynman rules are given in ref.\cite{pg2}. In the absence of
any fine cancellation, the dominant contribution
to loop corrected neutrino masses arises from $(i)$ and $(ii)$ above, when 
right sneutrinos are in the loop. This contribution
is proportional to squared mass difference between right
sneutrino scalar and pseudoscalar mass eigenstates \cite{Porod-Bartl,loop1,loop2,loop3}.

\vspace{0.2cm}
Using the dimensional reduction ($\overline{DR}$) scheme  in the 't-Hooft-Feynman
gauge ($\xi = 1$) one can write down the expression for one-loop corrected 
neutrino mass matrix in the basis, where the tree level mass matrix is diagonalized. 
This one-loop corrected neutrino mass matrix can be further diagonalized to 
get one-loop corrected mass eigenvalues.
In order to obtain the neutrino mixing matrix we rotate back
to the flavour basis using neutralino mixing matrix \cite{pg2}. In this way
we obtain the one loop corrected neutrino mass matrix in the flavour basis
and further apply the known methods of seesaw mechanism to get the one-loop
corrected mass eigenvalues $(m'_{\nu_i})$. The corresponding unitary matrix $U'_{PMNS}$,
which diagonalizes this matrix contain the 
one-loop corrected mixing angles \cite{pg2}.

The one-loop corrected neutrino mass matrix as shown in ref. \cite{pg2}
looks like

\beq
(M^{seesaw}_{loop-corrected})_{ij} = A_1 a_i a_j + A_2 c_i c_j + A_3 (a_i c_j + a_j c_i),
\label{one-loop corrected structure of neutralino mass matrix}
\eeq

where $A_i$s are functions of our model parameters and the Passarino-Veltman
functions $(B_0,B_1)$ \cite{Passarino-Veltman,Veltman-tHooft}. The form of 
the loop corrected mass matrix thus obtained is identical to the tree level one 
(see, eq.(\ref{mnuij-compact-recasted}))
with different coefficients $A_i$s in front arising because of the one-loop corrections.

\vspace{0.2cm}
The effect of one-loop corrections to neutrino masses and mixing varies with 
the neutrino mass hierarchy. We found it relatively easier to
accommodate the normal hierarchical spectrum of light neutrino masses compared to
inverted or quasi-degenerate spectrum when the effect 
of one-loop correction was taken into account \cite{pg2}. The allowed
region of tree level parameter space were observed to shrink severely
in the case of inverted hierarchy of light neutrino mass and thereby
only a very little window of allowed parameter space remains open with the inclusion
of loop-effect \cite{pg2}. The effect of one-loop correction produces
practically vanishing allowed region in parameter space for the quasi-degenerate 
scheme. However, we must emphasize here
that these conclusion are parameter dependent and the huge parameter
space of $\mu\nu$SSM always left us with enough room to observe some different
phenomena in a entirely different corner of parameter space.

\section{Decays of the LSP}
$\rpv$ in $\mu\nu$SSM leaves no room
for a stable LSP. The lightest SUSY particle in this model will eventually
decay into the SM particles making this model testable in collider experiment. 
It is an well-known feature of the SUSY models with
bilinear $\rpv$ to show nice correlations among neutrino oscillation parameters
and LSP decay patterns\cite{corr1,corr2,corr3}. In the $\mu\nu$SSM one
observes similar correlations among certain neutrino mixing angles with
ratio of decay branching ratios \cite{pg1,Porod-Bartl}. These correlations
vary with the concerned neutrino mass hierarchy and with the LSP nature \cite{pg1}.
In $\mu\nu$SSM, apart from a {\it{gaugino}} or {\it{higgsino}} LSP one can also
have a right-handed neutrino or {\it{singlino}} like LSP. This third possibility
is a special feature of $\mu\nu$SSM and is of great interest because of its direct relation 
with the seesaw scale. A proper detection may be followed by a faithful mass 
reconstruction tool to probe the right-handed neutrino mass
scale or the seesaw scale which is hitherto unseen for its gauge-singlet nature \cite{pg1}.
Decays of fermionic LSP can produce multiple leptons and jets 
in the final state decay products, for which a comparative analysis can be
performed for obtaining signatures of $\mu\nu$SSM at colliders \cite{Porod-Bartl}.
In $\mu\nu$SSM right-sneutrino is also an eligible 
candidate to be the LSP \cite{pg1}.

One more crucial experimental signature of $\mu\nu$SSM can come from the
study of displaced vertices. Depending on the LSP composition
the decay length can vary several orders of magnitude, which
is $\sim$ a few {\it{meters}} for a {\it{singlino}} like LSP. As
a corollary a dedicated investigation of the displaced vertices
can provide characteristic signature of $\mu\nu$SSM.

\ack
I would like to thank the Council of Scientific and Industrial 
Research, Govt. of India for  a Senior 
Research Fellowship. I gratefully acknowledge P. Dey, B. Mukhopadhyaya and S. Roy
for their collaborative supports. I also wish to thank the organizers
of the PASCOS 2010, ``16th International Symposium on particles, strings and cosmology''
for their kind hospitality. I would further like to thank my supervisors U. Chattopadhyay 
and S. Roy for providing financial assistance through their respective
research grants. Finally, I am grateful to Prof. K. Bhattacharya, Director, IACS for his kind
approval and for the encouragement I received from him.

\section*{References}
\bibliography{iopart-num}
\end{document}